\documentclass[10pt]{emulateapj}
\usepackage{amsmath}
\usepackage{graphicx}
\usepackage{natbib}
\usepackage{ulem}
\begin{document}

\title{Variability of Water and Oxygen Absorption Bands\\in the
Disk-Integrated Spectra of the Earth}

\author{
   Yuka Fujii\altaffilmark{1},
   Edwin L.~Turner\altaffilmark{2,3}, and
   Yasushi Suto\altaffilmark{1,2,4}.
}
\altaffiltext{1}{Department of Physics, The University of Tokyo,
Tokyo 113-0033, Japan}
\altaffiltext{2}{Department of Astrophysical Sciences, Princeton
University, Princeton, NJ 08544}
\altaffiltext{3}{Kavli Institute for the Physics and Mathematics of the
Universe, The University of Tokyo, Kashiwa 277-8568, Japan}
\altaffiltext{4}{Research Center for the Early Universe,
Graduate School of Science, The University of Tokyo, Tokyo 113-0033, Japan}
\email{yuka.fujii@utap.phys.s.u-tokyo.ac.jp}
\begin{abstract}
We study the variability of major atmospheric absorption features in the 
disk-integrated spectra of the Earth with future application to Earth-analogs in mind, concentrating on the diurnal timescale. We first analyze 
observations of the Earth provided by the EPOXI mission, and find 5-20\% 
fractional variation of the absorption depths of H$_2$O and O$_2$ bands, 
two molecules that have major signatures in the observed range. From a 
correlation analysis with the cloud map data from the Earth Observing 
Satellite (EOS), we find that their variation pattern is primarily due to 
the uneven cloud cover distribution.  In order to account for the observed 
variation quantitatively, we consider a simple {\it opaque cloud} model, 
which assumes that the clouds totally block the spectral influence of the 
atmosphere below the cloud layer, equivalent to assuming that the incident 
light is completely scattered at the cloud top level. The model is 
reasonably successful, and reproduces the EPOXI data from the pixel-level 
EOS cloud/water vapor data.  A difference in the diurnal variability 
patterns of H$_2$O and O$_2$ bands is ascribed to the differing vertical 
and horizontal distribution of those molecular species in the atmosphere. 
On the Earth, the inhomogeneous distribution of atmospheric water vapor is 
due to the existence of its exchange with liquid and solid phases of 
H$_2$O on the planet's surface on a timescale short compared to 
atmospheric mixing times. If such differences in variability patterns were 
detected in spectra of Earth-analogs, it would provide the information on 
the inhomogeneous composition of their atmospheres. 
\end{abstract}
\keywords{Earth -- scattering -- techniques: spectroscopic}

\section{Introduction}
\label{s:intro}

Determining the nature of the atmospheres and surfaces of exoplanets is of
primary importance in probing not only their formation history but also in 
identifying
possible signatures of life. While it is very challenging, and perhaps 
only feasible through direct
photometric/spectroscopic observations of the planetary light resolved
from that of the host star, there are several proposals for eventual
astrobiological investigations of potentially habitable, rocky exoplanets 
\citep[e.g.,][]{levine2009,savransky2010,matsuo2010}.

The available exoplanetary light would be
disk-integrated, i.e., that of a point source without any
spatial resolution. Deciphering the exoplanetary light properly,
therefore, is inevitably a highly difficult task, in particular for
those planets with diverse surface types and atmospheres like our own
Earth.

Indeed, habitable planets are likely to exhibit a variety of complex
patterns of their surfaces and atmospheres, intrinsically dependent on
their climatology and geology. According to the global water cycle,
liquid water on the planetary surface vaporizes, forms clouds, is
carried by atmospheric circulation, and precipitates as
rainfall/snowfall. Depending on the total amount of water and the
atmospheric circulation pattern, the surface of the habitable planets
may be partially covered by ocean \citep[e.g.][]{abe2011}, and be
observed only through atmospheres with highly inhomogeneous and variable 
cloud
cover patterns.

Given these complexities, techniques to properly decipher the
disk-integrated light of exoplanets need to be developed.  Towards that
goal, the time variation of planetary light due to spin rotation and
orbital revolution is a powerful tool, and several authors have computed 
the
expected variation patterns and proposed the reconstruction methods
\citep{ford2001,tinetti2006a,tinetti2006b,cowan2009,oakley2009,
kawahara2010,robinson2010,cowan2011,robinson2011,kawahara2011,
fujii2010,fujii2011,fujii2012,sanroma2012}.

The variation of the continuum level in the visible to near-infrared (NIR) 
range mainly reflects the distribution of landmass, ocean, cloud cover, 
and possibly vegetation.
The peak-to-trough diurnal variability for 0.1$\mu $m-wide photometry of 
the
Earth in the visible/NIR range is found to be 10-30\% \citep{livengood2011}.
Thermal emission of the Earth also shows a few percent of diurnal 
variation in the mid-infrared \citep{gomezleal2012}, which primarily 
originates from the uneven cloud cover and humidity.

In addition to the light-curve in broad-band photometry mentioned above,
molecular absorption depths exhibit diurnal
variation. Since atmospheric absorption depths of molecules are
determined by the column density of the corresponding molecules along
the optical path, they are sensitive to the presence of highly
reflective cloud cover in the atmosphere that effectively blocks
the spectral influence of molecules in the lower atmosphere. For idealized
uniformly mixed atmospheres, the diurnal variation of molecular
absorption features should be strongly linked to the spatial distribution 
of clouds.

In reality, the distribution of molecules itself is not entirely uniform 
and may be
time-dependent. Moreover, the spatial and time variations are not the same 
for different
molecular species comprising the atmosphere.
In particular, the local column density of water vapor is known to vary 
widely on a timescale of $\sim $ 1 hr \citep{blake2011}, while other 
molecules, such as N$_2$ and O$_2$, are well-mixed in the troposphere and 
thus
show no significant variations on short timescales.

This paper extends our previous work on photometric light-curves in
visible/NIR bands \citep{fujii2010,fujii2011}, and examines the diurnal
variation of molecular absorption signatures in the disk-integrated
reflection spectra of the Earth with future application to Earth-analogs 
in mind\footnote{The term ``Earth-like'' has been used extensively in the
literature, often without any precise definition.  In this paper, the 
term ``Earth-analogs'' is used to 
refer to rocky planets that resemble the Earth in having surface temperatures, 
obliquities, continents, oceans and atmospheres sufficiently like the Earth's
to give them similar global hydrologies.}.
We focus on the absorption bands of
H$_2$O at 1.13$\mu $m and O$_2$ at 1.27$\mu $m, which are among the most
prominent absorption bands in NIR and are an indicator of habitability and a 
biosignature molecule, respectively.
We first analyze data
from space-based NIR spectroscopy of the Earth by NASA's EPOXI
mission. Then we consider a simple model (referred to as {\it opaque
cloud model}) that reasonably reproduces the observed variation pattern,
if the cloud pattern data are provided separately. Our model implies that 
the different behavior of water vapor
and oxygen in their absorption band variations can be ascribed to their
intrinsically different spatial distribution patterns, 
rather than
to the common cloud coverage. We discuss how this difference can be used
to probe signatures of surface and atmospheric inhomogeneity of
exoplanets with next-generation direct imaging.

The organization of this paper is as follows. Section \ref{s:data}
introduces the EPOXI data used in this paper and analyses the
correlation between absorption depths and pixel-to-pixel climatological
data obtained with Earth Observing Satellites.  Section \ref{s:model}
describes our model to reproduce the variation pattern of absorption
depths and compares the simulation results with observation.  Section
\ref{s:dis} further discusses the different behavior between H$_2$O
variation and O$_2$ variation.  Finally, Section \ref{s:sum} draws our
main conclusions and discusses the implication for future observation of
exoplanets. An analysis of CO$_2$, a molecule with properties intermediate
between those of H$_2$O and O$_2$ in some ways, is presented in Appendix \ref{ap:CO2}.

\section{Analysis of EPOXI Data}
\label{s:data}

\subsection{EPOXI observation}
\label{ss:EPOXI}

\begin{figure}[!h]
     \begin{center}
       \includegraphics[width=0.9\hsize]{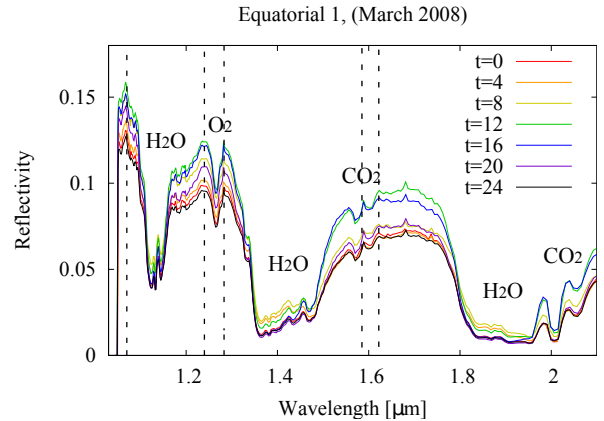}
     \end{center}
\caption{Examples of NIR reflection spectra of the Earth observed by
EPOXI in March of 2008 (Earth 1: equinox) 
\citep{livengood2011,robinson2011} . Different lines represent different 
short exposures obtained at 2-hour intervals and
labeled by the time $t$[hr] from the start of the observations on that 
date (see Figure \ref{fig:diurnalvar}). Three vertical dashed lines indicate the wavelength 
ranges that we adopt to
compute the equivalent widths (H$_2$O: 1.07-1.24$\mu $m, O$_2$: 1.24$\mu 
$m-1.283$\mu $m, CO$_2$: 1.59-1.62$\mu $m). } \label{fig:spec}
\end{figure}

\begin{figure*}[!hbt]
   \begin{minipage}{0.24\hsize}
     \begin{center}
       \includegraphics[width=\hsize]{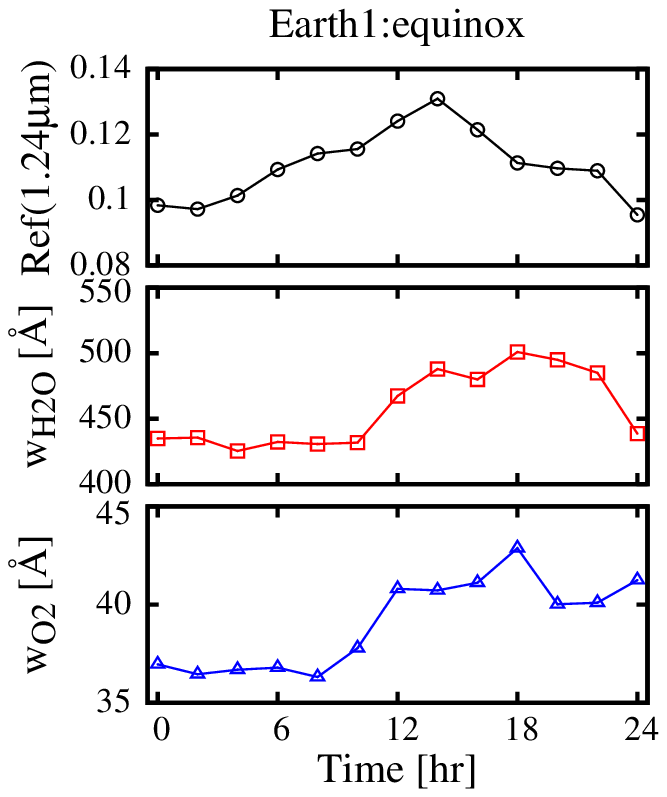}
     \end{center}
\end{minipage}
   \begin{minipage}{0.24\hsize}
     \begin{center}
       \includegraphics[width=\hsize]{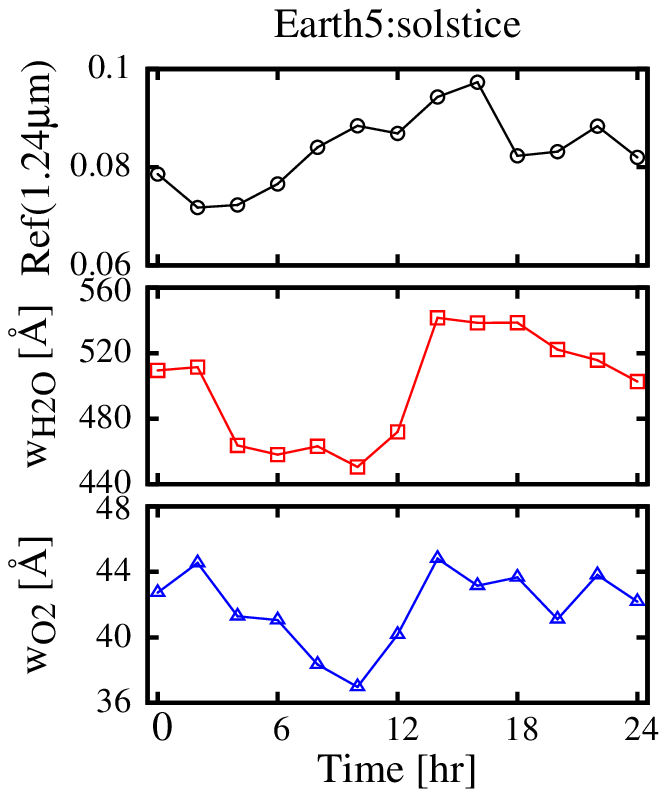}
     \end{center}
\end{minipage}
   \begin{minipage}{0.24\hsize}
     \begin{center}
       \includegraphics[width=\hsize]{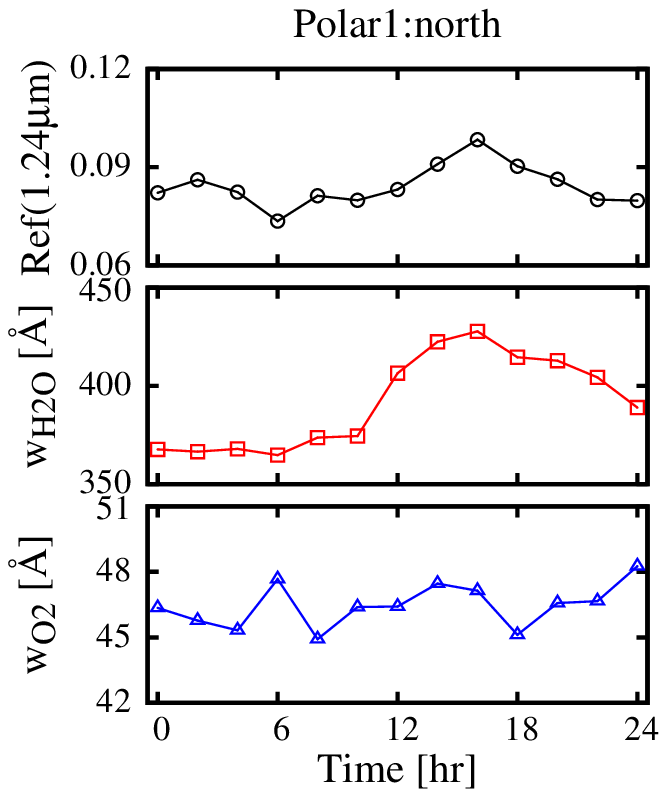}
     \end{center}
\end{minipage}
   \begin{minipage}{0.24\hsize}
     \begin{center}
       \includegraphics[width=\hsize]{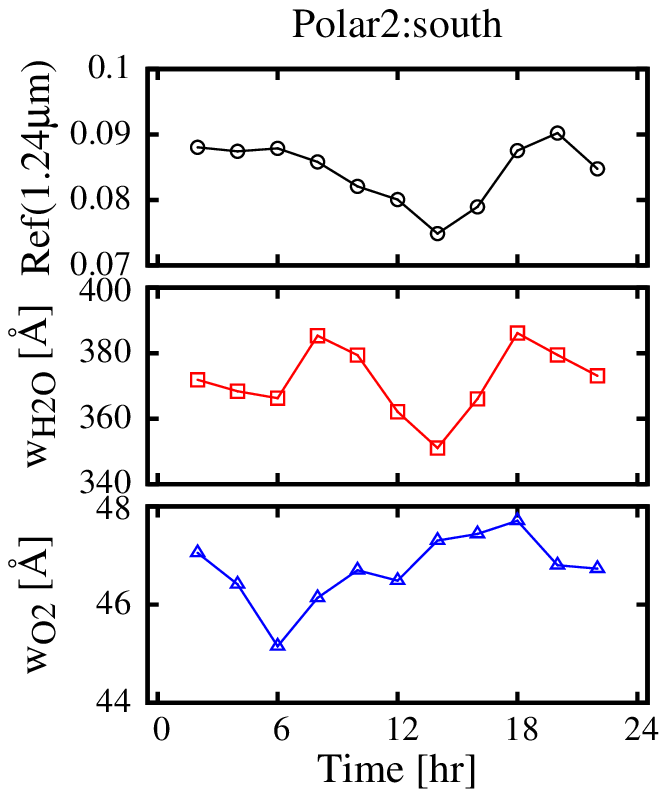}
     \end{center}
\end{minipage}
   \begin{minipage}{0.24\hsize}
     \begin{center}
       \includegraphics[width=\hsize]{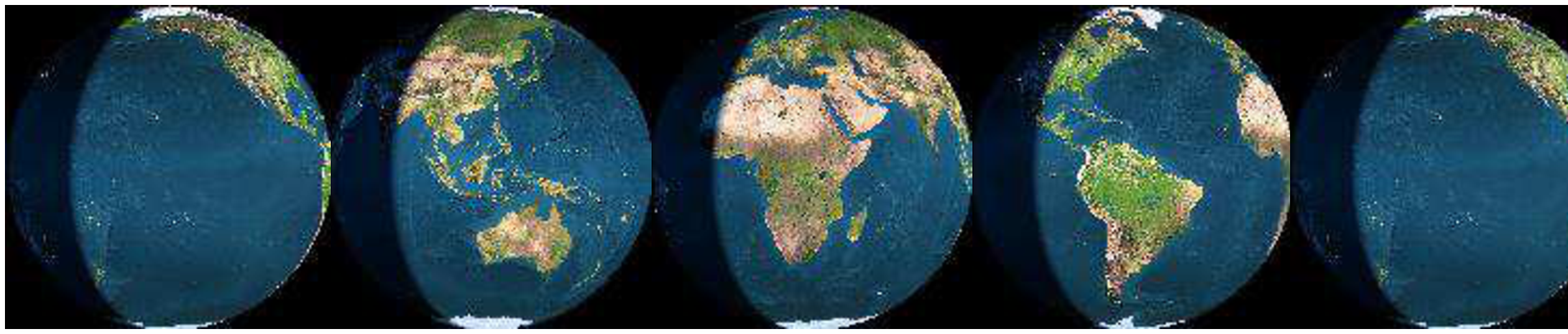}
     \end{center}
\end{minipage}
   \begin{minipage}{0.24\hsize}
     \begin{center}
       \includegraphics[width=\hsize]{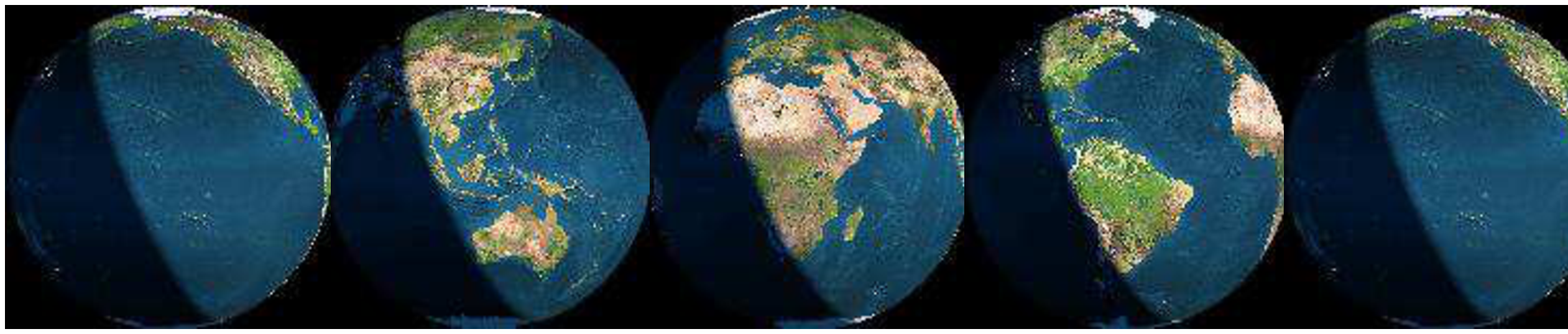}
     \end{center}
\end{minipage}
   \begin{minipage}{0.24\hsize}
     \begin{center}
       \includegraphics[width=\hsize]{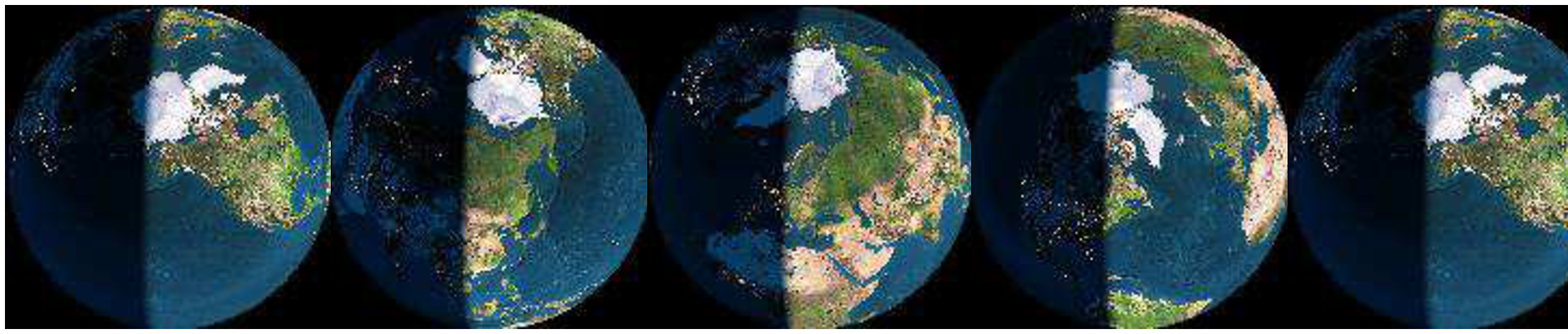}
     \end{center}
\end{minipage}
   \begin{minipage}{0.24\hsize}
     \begin{center}
       \includegraphics[width=\hsize]{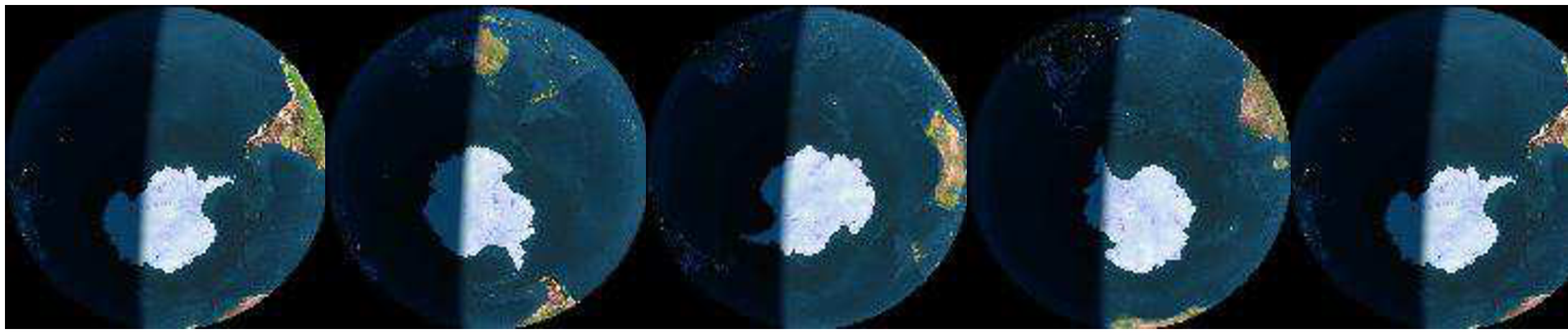}
     \end{center}
\end{minipage}
\caption{Diurnal variations of continuum level (top), equivalent width
of H$_2$O absorption centered at 1.13$\mu $m (middle), and that of O$_2$
centered at 1.27$\mu $m (bottom). Symbols indicate observed data points 
and the connecting lines are drawn only to guide the eyes. Snapshots at
corresponding times were generated at http://www.fourmilab.ch/cgi-bin/uncgi/Earth are attached. }
\label{fig:diurnalvar}
\end{figure*}

\begin{table*}[!htdp]
\begin{center}
\caption{Diurnal Variation of Continuum Level, Equivalent Width of H$_2$O 
at 1.13$\mu $m and that of O$_2$ at 1.27$\mu $m. } \label{tab:var}
\begin{tabular}{l|cc|cc|cc} \hline \hline
& \multicolumn{2}{c|}{Reflectivity at 1.24$\mu $m} & 
\multicolumn{2}{c|}{H$_2$O at 1.13$\mu $m}& \multicolumn{2}{c}{O$_2$ at 
1.27$\mu $m} \\
& ave.        & (max-min)/ave & ave.[\AA]        & (max-min)/ave & ave.[\AA] & 
(max-min)/ave \\ \hline
Earth1:equinox (Mar.2008) &0.111        &32.1\%        &457 &16.5\%& 39&16.8\%\\
Earth5:solstice (Jun.2008) &0.084        &30.6\%        &499 &18.2\%&42 &18.7\%\\
Polar1:north (Mar.2009)        &0.084        &29.5\%        &392 &16.1\%& 46&7.2\%\\
Polar2:south (Oct.2009)        &0.084        &18.2\%        &372 &9.4\%& 47&5.5\%\\ \hline
\end{tabular}
\end{center}
\end{table*}%

The EPOXI\footnote{The Deep Impact flyby spacecraft for the Extrasolar
Planetary Observation and Characterization investigation (EPOCh) and the
Deep Impact eXtended Investigation (DIXI)} mission \citep{livengood2011}
performed photometric and spectroscopic monitoring of the Earth
and the Moon from space as a benchmark for future characterization of
terrestrial exoplanets.  A part of the observations were devoted to
spectroscopy of the disk-integrated scattered light of the Earth over
the wavelength of 1.10-4.54$\mu $m.  These observations were carried out
on 2008 March 18-19, 2008 June 4-5, 2009 March 27-28, and 2009 October
4-5, with 12 exposures every two hours of the day (integration time for
each exposure is less than 2 sec).
The sub-observer latitudes (the latitude of the intersection between the 
Earth's surface and the line connecting the center of the Earth and the 
detector) are 1$^{\circ }$.7 N, 0$^{\circ }$.3 N, 61$^{\circ }$.7 N, and 
$73^{\circ }$.8 S,
respectively.
Following \citet{cowan2011}, we henceforth refer to these 4 observations 
as Earth1:equinox, Earth5:solstice, Polar1:north, and Polar2:south.

Figure \ref{fig:spec} displays the 1-2$\mu $m portion of the observed
reflection spectra of the Earth in Earth1.  The broad absorption
features at 1.08-1.18, 1.30-1.53, and 1.75-1.99$\mu $m are mostly due to
H$_2$O, and the narrower feature around 1.27$\mu $m is due to O$_2$ plus
oxygen collision complexes O$_2\cdot $O$_2$ and O$_2\cdot $N$_2$
\citep[][and references therein]{palle2009}.  Absorptions at 1.6$\mu $m
and 2.0$\mu $m are signatures of CO$_2$ \citep[e.g.][]{robinson2011}.

We measure the equivalent widths of H$_2$O ($w_{\rm H_2O}$) and O$_2$
($w_{\rm O_2}$) for each exposure.  For $w_{\rm H_2O}$, we focus on the
spectral features centered at $\sim 1.13\mu$m and consider the
absorption from 1.07$\mu $m to 1.24$\mu $m.  For $w_{\rm O_2}$, we use
the spectral features centered at $\sim 1.27\mu$m and consider the
absorption from 1.24$\mu $m to 1.283$\mu $m.
In each case, the continuum line is assumed to connect the data points at 
both boundaries linearly.
Additionally, we consider the variation of reflectivity at 1.24$\mu $m as 
a measure of the continuum level.

Figure \ref{fig:diurnalvar} shows the diurnal fluctuations of $w_{\rm 
H_2O}$
and $w_{\rm O_2}$, as well as the variation of the reference continuum
level (1.24 $\mu $m).  Table \ref{tab:var} summarizes the average and
the fractional variation amplitudes of continuum level, $w_{\rm H_2O}$ and 
$w_{\rm O_2}$.  The variation amplitude defined by 
(maximum-minimum)/average is typically
5-20\%. Figure \ref{fig:diurnalvar} clearly indicates that the diurnal 
patterns
of $w_{\rm H_2O}$ and $w_{\rm O_2}$ significantly differ from that of
the continuum level which primarily traces the continental distribution
on the Earth \citep[see e.g.][]{cowan2009,cowan2011,fujii2011}.
Although not shown here in detail, we also confirmed that absorption 
features of H$_2$O
at other wavelengths (centered at $\sim 1.4\mu $m and $\sim 1.85\mu $m)
exhibit variation patterns matching those at the 1.13$\mu $m band.  For 
equatorial observations,
the general trend that the depth is weaker in the first half of the day
and becomes stronger in the second half is evident, and is shared by
both H$_2$O and O$_2$.  Referring to the snapshots shown in the bottom
panels in Figure \ref{fig:diurnalvar}, the weaker absorption corresponds
to the time when the Indonesia, a persistently cloudy region, dominates 
the field-of-view
  \citep[see also][]{gomezleal2012}. This already demonstrates the strong 
relation between the absorption bands and cloud
cover.  In the next subsection, we investigate this connection in more
detail.

\subsection{Correlation between diurnal variabilities
and ocean/cloud parameters} \label{ss:corr}

As mentioned in Section \ref{ss:EPOXI}, the variation pattern in Figure
\ref{fig:diurnalvar} is likely correlated with the extent and nature of 
cloud
cover.  In order to confirm the correlation between the absorption
features and clouds, we collect daily global maps of atmospheric
parameters for the corresponding days from Terra/MODIS Atmosphere Level 3
Product, which is available online\footnote{ 
http://ladsweb.nascom.nasa.gov/
}. Each global map is derived on a pixel-to-pixel basis from the Remote
Sensing data obtained with MOderate Resolution Imaging Spectroradiometer
(MODIS) onboard the Earth Observing Satellites {\it Terra} and {\it
Aqua} \citep{MODIS}.  Among the various parameters reported, we choose
four which we suspect will influence the diurnal absorption variations 
strongly,
including {\it Atmospheric Water Vapor Mean}, {\it Cloud Top Pressure
Mean} ($P_{\rm ctp}$), {\it Cloud Fraction Mean} ($f_{\rm cld}$), and
{\it Cloud Optical Thickness Combined Mean} ($\tau _{\rm cld}$).  We
further add the ocean fraction as a fifth parameter.

We compute the {\it weighted} average of each parameter for each EPOXI
exposure.  We adopt the geometric weight as ${\rm Max} \{\mu _0 \mu _1 ,
0\}$, where $\mu _0$/$\mu _1$ denote the cosine between the normal
direction of the surface and the direction toward the Sun/observer.
We adopt this factor because each surface pixel would
contribute to the total reflectivity with that weight if the scattering
were isotropic (Lambert's Law) \citep[e.g.][]{lester1979}; while this
is not strictly the case, it is a reasonable approximation for these 
purposes.

\begin{figure}[!h]
     \begin{center}
       \includegraphics[width=0.9\hsize]{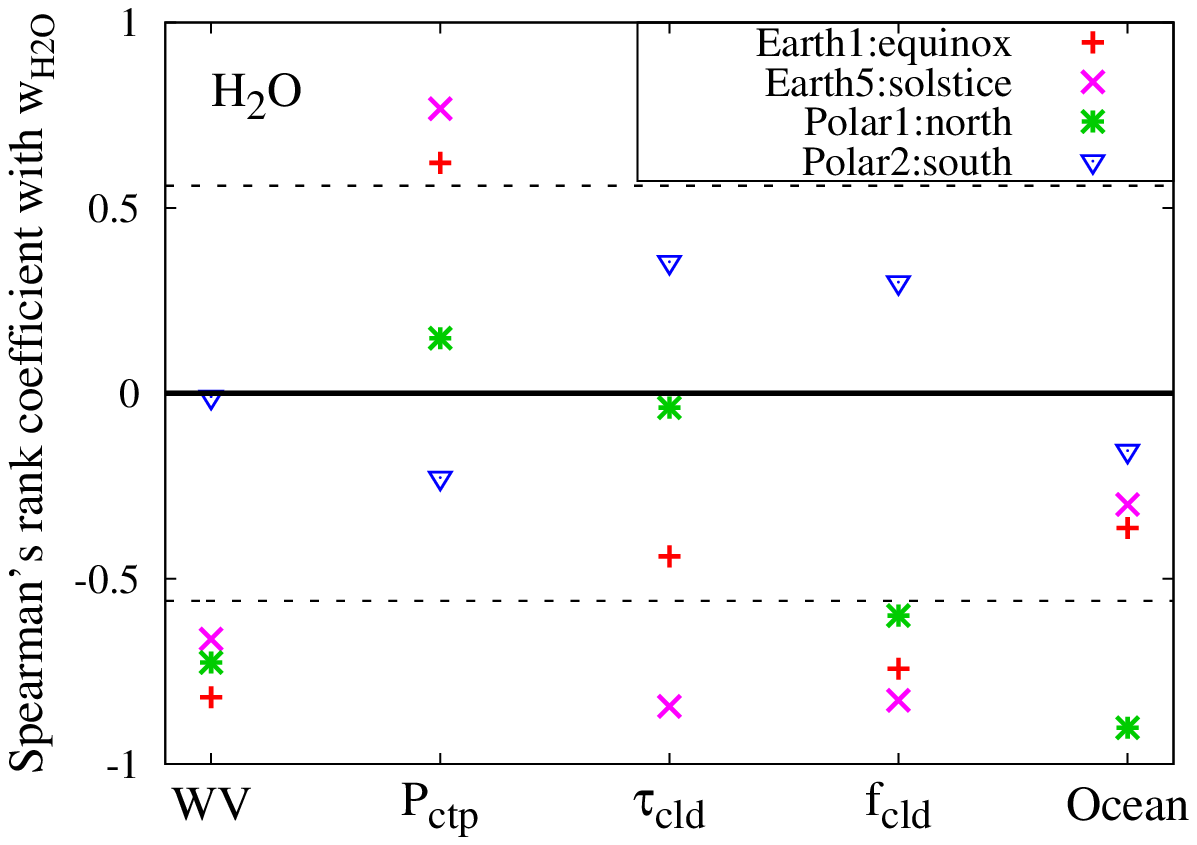}
       \includegraphics[width=0.9\hsize]{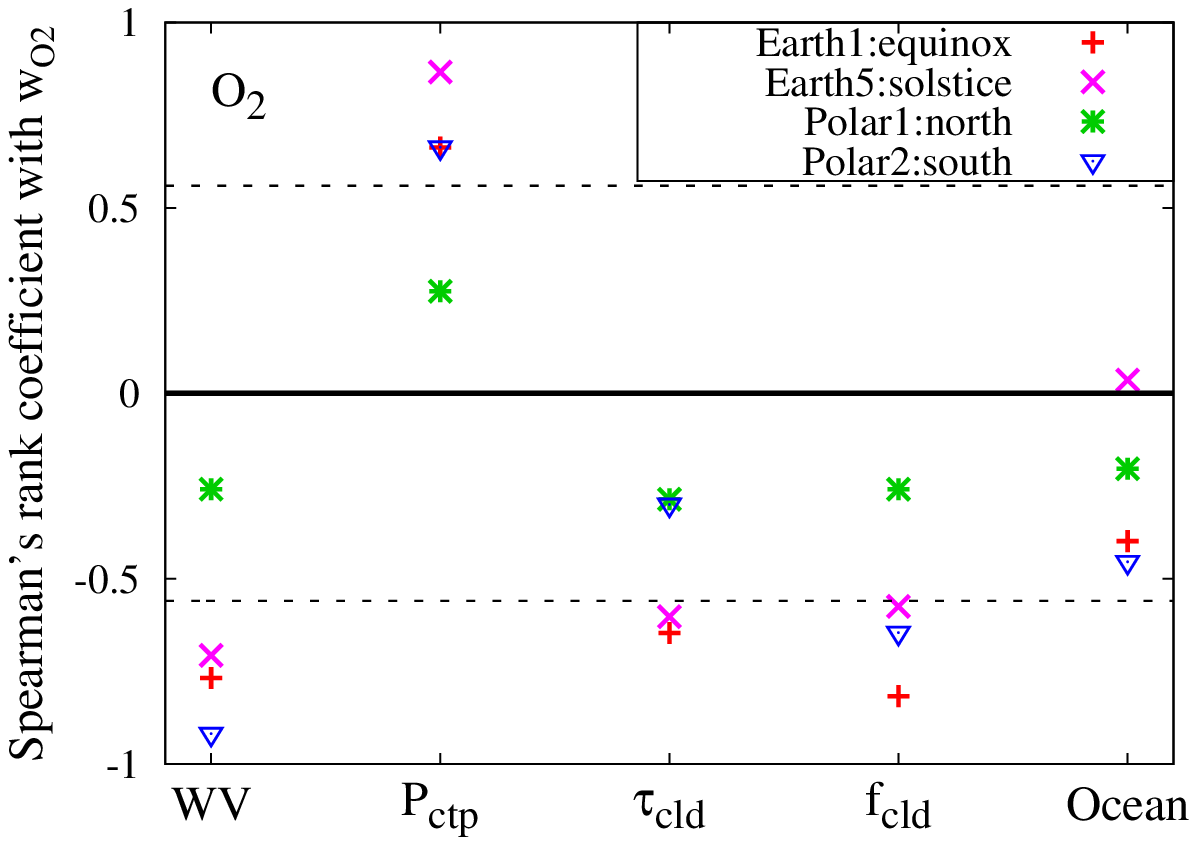}
     \end{center}
\caption{Spearman's rank correlation coefficient between absorption
depth (left:H$_2$O, right:O$_2$) and Atmospheric Water Vapor (WV), Cloud
Top Pressure ($P_{\rm ctp}$), Cloud Optical Thickness ($\tau _{\rm
ftp}$), Cloud Cover Fraction ($f_{\rm cld}$), and Ocean Fraction
(Ocean), respectively. Horizontal dotted lines show the significance
level of 0.05 for 13 samples (\# of exposures per observation). }
\label{fig:corr}
\end{figure}

Figure \ref{fig:corr} displays Spearman's rank correlation coefficients
between absorption depths of H$_2$O (left) and O$_2$ (right) measured by
EPOXI observation and parameters taken from MODIS.  Spearman's rank
coefficient $r_s$ is calculated by converting each value $x_i$
($i$:index for exposures in one series of observation) to the rank
$X_i$.  If $x_j$ is the $n$-th largest value of all possible values of
$x_i$, the rank is defined as $X_j=n$. Then we calculate the correlation
coefficient of the ranks of two different parameters $x_i$ and $y_i$ as
\begin{eqnarray}
r_s &=& \frac{\displaystyle \sum_i(X_i-\bar X)(Y_i-\bar Y)}
{\sqrt{\displaystyle \sum_i {(X_i-\bar X)^2}}{\sqrt{\displaystyle \sum_i 
{(Y_i-\bar Y)^2}}}}, \\
\bar X &=& \sum _i X_i, \qquad \bar Y = \sum_i Y_i .
\end{eqnarray}
Since the chosen parameters do not necessarily follow a gaussian
distribution, we adopt Spearman's rank correlation coefficients
instead of Pearson's correlation coefficients.

In most cases, the absorption depths are positively correlated with the
cloud top pressure (or equivalently, anti-correlated with the cloud
altitude), while anti-correlated with both cloud optical thickness and
the cloud cover fraction.  These trends are plausibly understood as
follows:  For cloud top pressure, the incident light is scattered at the
altitude of the upper cloud layer without absorption by molecules in
the atmosphere below the cloud. Since molecules are abundant in the
lower atmosphere, the higher cloud altitude reduces the absorption depth
more effectively.  As for cloud optical thickness, thicker clouds scatter
the incident light more completely and thus reduce absorptions below
that layer.  Additionally, the higher reflectivity of the thicker cloud
increases the contribution of the region to the disk-integrated spectra.
Similarly, the absorption depths become weaker as the cloud cover
fraction increases.

Although more weakly, the absorption depths appear to be slightly
anti-correlated with the ocean fraction. The lower reflectivity
of ocean reduces the absorption depths of the disk-integrated
spectra\footnote{The light reflected from ocean in principle contains
the signature of {\it liquid} water \citep{palmer1974}, which is shifted
toward longer wavelength and less prominent compared to the signature of
water {\it vapor} we consider in this paper.}.

We should
emphasize here, however, that the above interpretation requires some
caution because the five parameters we adopted are not necessarily
independent.  In the equatorial region, for instance, there exists a
positive correlation between cloud optical thickness and cloud top
altitude.  There is also generally positive correlation between cloud 
cover and
atmospheric water vapor, likely leading to the
counter-intuitive anti-correlation between the atmospheric water vapor
and the absorption depths.

Figure \ref{fig:corr} also shows that the correlation of the H$_2$O band
with clouds varies more significantly than that of O$_2$, depending on
the location of the sub-observer's latitude. In particular, the
correlation of H$_2$O for Polar2 observation even changes its sign
compared to the other three cases. This is not the case for O$_2$.
We will discuss this behavior in detail in Section \ref{s:dis}.

\section{A Simple Opaque Cloud Model}
\label{s:model}

In the previous section, we discussed the influence of clouds on
absorption features by considering the correlation between the
absorption depths and individual cloud/surface parameter.  In this
section, we try to interpret the observed variation in absorption depths
more quantitatively. For that purpose, we consider a simple {\it opaque 
cloud} model \citep[see also e.g.][]{harrison1988}, which assumes that the 
cloud cover totally blocks the influence of the lower atmosphere and that 
the incident light is completely scattered at the cloud top.

\subsection{Model Description}
\label{ss:model}

The equivalent width of an atmospheric absorption band in the
disk-integrated scattered light of the planet can be expressed
as
\begin{equation}
\label{eq:w_theory_org}
w _{\rm model}= \int _{\lambda _0 - \Delta \lambda }^{\lambda _0 +
  \Delta \lambda } d\lambda ~
\frac{\int_{S_{\rm IV}} c_{\lambda }(\theta, \phi )
\left[ 1 - e^{-\tau _{\lambda }} \right] g(\theta , \phi)\;d\Omega}
{\int_{S_{\rm IV}}c_{\lambda }(\theta, \phi ) g(\theta , \phi)\;d\Omega }, 
\end{equation}
\begin{equation}
d\Omega \equiv  \sin \theta d\theta d \phi,
\qquad g (\theta , \phi) \equiv  \mu _0 \mu _1,
\end{equation}
where $(\theta, \phi)$ represents the polar-coordinate of the surface
point on the planet, $\mu _0=\mu _0(\theta, \phi)$ and $\mu _1=\mu
_1(\theta, \phi)$ are the directional cosines between the normal
direction of the surface point and the direction toward the Sun and the
observer, respectively.
The wavelength-dependent continuum level (i.e., reflectivity) and optical
depth are denoted by $c_{\lambda }(\theta, \phi)$ and $\tau _{\lambda
}$. The interval of integration over $\lambda $ is centered at the
center of the absorption band, $\lambda _0$.  The integral over solid
angle of planetary surface is performed over the illuminated and visible
portion, $S_{\rm IV}$.  The above expression is based on the
approximations that 1) the lower boundary scatters the light according
to the Lambert Law, i.e. the scattered intensity is independent of the
emergent direction, and 2) scattering by the atmosphere is negligible.

We further neglect the wavelength-dependence of the continuum level, and
assume that the wavelength dependence of $c_{\lambda }(\theta, \phi)$
can be factored out, i.e., $c_{\lambda }(\theta, \phi)
= h(\lambda )c(\theta, \phi)$ with $h(\lambda)$ being a function of
wavelength alone.
Then, equation (\ref{eq:w_theory_org}) may be approximated as
\begin{eqnarray}
\label{eq:w_theory}
w _{\rm model}&=&
\frac{\int_{S_{\rm IV}} c(\theta, \phi )
W(N; \theta , \phi) g (\theta , \phi) d\Omega}
{\int_{S_{\rm IV}} c(\theta, \phi ) g (\theta , \phi) d\Omega} , \\
\label{eq:W}
W(N; \theta , \phi)&\equiv & \int d\lambda \;
\left( 1 - \exp \{-\tau _{\lambda }(N; \theta , \phi)\} \right) , \\
\label{eq:tau}
\tau _{\lambda }(N; \theta , \phi) &=& \int_{z_0}^{\infty} n(z) 
\sigma_\lambda \left( \frac{1}{\mu_0} + \frac{1}{\mu _1} \right) dz \notag \\
&\sim &\left( \frac{1}{\mu_0} + \frac{1}{\mu _1} \right) \frac{\int n (z) dz 
\cdot \int n_0 (z) \sigma _{\lambda }dz}{\int n_0 (z) dz} \notag \\
&=& \frac{N_{\rm eff}}{N_0} \tau _{\lambda, 0}\; , \label{eq:taulambda}\\
N_{\rm eff} &\equiv & \left( \frac{1}{\mu_0} + \frac{1}{\mu _1} 
\right)\int_{z_0}^{\infty } n (z) dz \; ,
\end{eqnarray}
where $W(N;\theta,\phi)$ is the equivalent width at each surface patch,
$\sigma_\lambda$ is the absorption coefficient, $N_0$ is the total column 
density of molecules under the canonical atmospheric model, $N_{\rm eff}$ 
is the effective column density obtained by integrating $n(z)$ along the 
optical path from the top of atmosphere down to the boundary at altitude 
$z=z_0$, and $\tau _{\lambda ,0}$ is the absorption depth of the canonical 
model.
Equation (\ref{eq:taulambda}) is only approximately valid because in 
reality the absorption coefficient $\sigma _{\lambda }$ depends on the T-P 
profile of atmosphere. Nevertheless, we employ this approximation for the 
sake of simplicity.

In practice, we divide the planetary surface into $2^{\circ }\times
2^{\circ }$ pixels, and treat the cloudless and cloudy portions
in each pixel separately for convenience, as described below.  Then,
equation (\ref{eq:w_theory}) is discretized as
\begin{equation}
\label{eq:Dtotal}
w _{\rm model} =
\frac{\displaystyle \sum _i \left\{ (1-f_{\rm cld}^i)\,W_{\rm nocld}^i 
c_{\rm nocld}^i
+ f_{\rm cld}^i\, W_{\rm cld} ^i c_{\rm cld}^i\right\}
g^i \delta \Omega^i}
{\displaystyle \sum _i \left\{ (1-f_{\rm cld}^i)\,c_{\rm nocld}^i
+ f_{\rm cld}^i\,c_{\rm cld}^i\right\} g^i \delta \Omega ^i} , 
\end{equation}
where $i$ is the index for the surface pixels, $f _{\rm cld}$ is the cloud
cover fraction, and the suffix $_{\rm nocld}$/$_{\rm cld}$ indicates the
value at the cloudless/cloudy portion.  The position-dependent
parameters in equation (\ref{eq:Dtotal}) are determined on the basis of
daily global maps which are extracted from the same MODIS data product 
used
in Section \ref{ss:corr}.  For instance, the value for $f_{\rm cld}$ is
adopted from {\it Cloud Fraction Mean}.
Input data for other parameters are described below.

The effective column density $N_{\rm eff}$ of H$_2$O at each patch is 
determined by {\it Atmospheric Water Vapor Mean} ($w$) and {\it Cloud Top 
Pressure Mean} ($P_{\rm ctp}$) as well as the geometric factors $\mu _0$ 
and $\mu_1$.
For the cloudless portion, $N_{\rm eff}$ is identical to
$(1/\mu_0 + 1/\mu_1)w$ because we can safely set $z_0 = 0$.  For cloudy
portions of a pixel, however, we need to {\it define} the boundary height 
$z_0$ in equation (\ref{eq:tau}) that is essentially the layer below which 
the
the molecule in question does not influence the emergent spectra. In the
present {\it opaque cloud} model, we assume that the synthetic
absorption depth is completely unaffected by the atmosphere below the
cloud top altitude $z_{\rm ctp}$, and thus set $z_0=z_{\rm
ctp}$. According to this opaque cloud assumption, $N$ is determined once
the vertical profile of water vapor and the altitude of the cloud top
layer are given.  In what follows, we simply assume that the vertical
profile of water vapor is proportional to that of the US Standard Model,
and determine the proportionality coefficient so that the total column
density matches the value of $w$.  The altitude of the cloud top layer
is derived from $P_{\rm ctp}$ assuming the US Standard
Temperature-Pressure (T-P) profile\footnote{T-P profiles are not 
significantly different in different atmospheric models.}.
The total column density $N_{\rm eff}$ for O$_2$ is estimated in the same 
way except that we neglect the horizontal inhomogeneity, and determine it 
only from $P_{\rm ctp}$ and the geometric factor.

\begin{figure}[!h]
     \begin{center}
       \includegraphics[width=0.9\hsize]{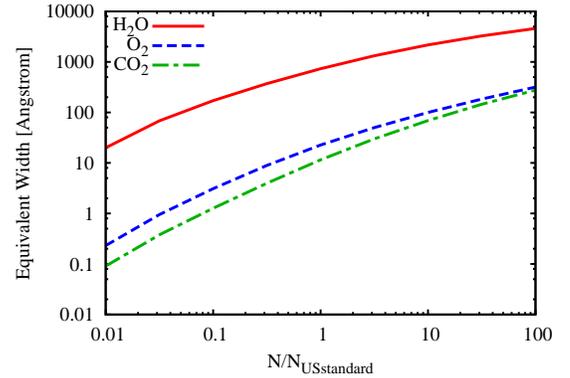}
     \end{center}
\caption{Curve of growth for the equivalent width of the water vapor 
absorption band at 1.13$\mu  $m-1.24$\mu $m, oxygen at 1.24$\mu 
$m-1.283$\mu $m, and carbon dioxide at 1588$\mu $m-1.622$\mu $m calculated 
by {\it lbl2od} with {\it GEISA 2011} database under the no-cloud 
condition with surface albedo 0.3. }
\label{fig:cog}
\end{figure}

The remaining task is to compute the equivalent width $W$ as a function of 
the effective column density. For this purpose, we compute line-by-line 
optical depths $\tau_{\lambda ,0}$ under the US Standard Atmosphere with the 
{\it GEISA 2011} molecule spectroscopic 
database\footnote{http://ether.ipsl.jussieu.fr/etherTypo/?id=1293} and the 
public code {\it 
lbl2od}\footnote{http://www.libradtran.org/doku.php?id=lbl2od}. Then we 
calculate $\int _{\lambda _0 - \Delta \lambda }^{\lambda_0 + \Delta 
\lambda } e^{-\kappa \tau_{\lambda ,0}} d\lambda $ with varying $\kappa $. 
Figure \ref{fig:cog} shows the resultant curve of growth of the equivalent 
widths of H$_2$O measured at $1.06-1.24\mu $m, O$_2$ at $1.235-1.310\mu $m as 
well as CO$_2$ at $1.560-1.630\mu m$ used in Appendix \ref{ap:CO2}. Based 
on these calculations, the normalized effective column density of each 
patch, $N^i_{\rm eff}/N_{\rm USstandard}$, is translated into the 
equivalent width $W^i$ by equating $N^i_{\rm eff}/N_{\rm USstandard}$ to 
$\kappa $ in Figure \ref{fig:cog}. 

Finally, the continuum level $c$ in equation (\ref{eq:Dtotal}) is
estimated using the 2-stream approximation \citep[e.g.][]{liou1980}.
We consider a non-absorbing atmosphere with optical thickness $\tau _{\rm 
cld}$ and asymmetry factor $\beta $. Denoting the reflectivity at the 
surface by $r_g$, the net reflectivity at the top of the atmosphere is:
\begin{equation}
\label{eq:2-stream}
c(\tau _{\rm cld}) = \alpha
\left[ 1 -  \frac{1-r_g}{1 + (\sqrt{3}/2)(1-r_g)(1-\beta )\tau _{\rm
  cld}}\right],
\end{equation}
where we adopt the optical thickness of the cloud ({\it
Cloud Optical Thickness Combined Mean}) for $\tau _{\rm cld}$.  The 
typical value for $\beta $ of Earth's clouds at the relevant wavelengths 
is $\beta=0.85$ \citep[e.g.][]{liou1980}.
The overall scaling factor $\alpha$ is introduced here to empirically incorporate 
the anisotropic scattering due to the bulk cloud
cover. We determine the value of $\alpha$ by hand so as to match the 
disk-averaged continuum level to the observed data.

According to the procedures described above, we compute $W^i$ and $c^i$ at 
each pixel, and then obtain the predicted
absorption depth in the disk-averaged spectra of the Earth from equation
(\ref{eq:Dtotal}).
The model prediction is compared with the EPOXI data in Section 
\ref{ss:comp}.

\subsection{Comparison of the opaque cloud model and the EPOXI Data}
\label{ss:comp}

\begin{figure*}[!bt]
   \begin{minipage}{0.33\hsize}
     \begin{center}
       \includegraphics[width=\hsize]{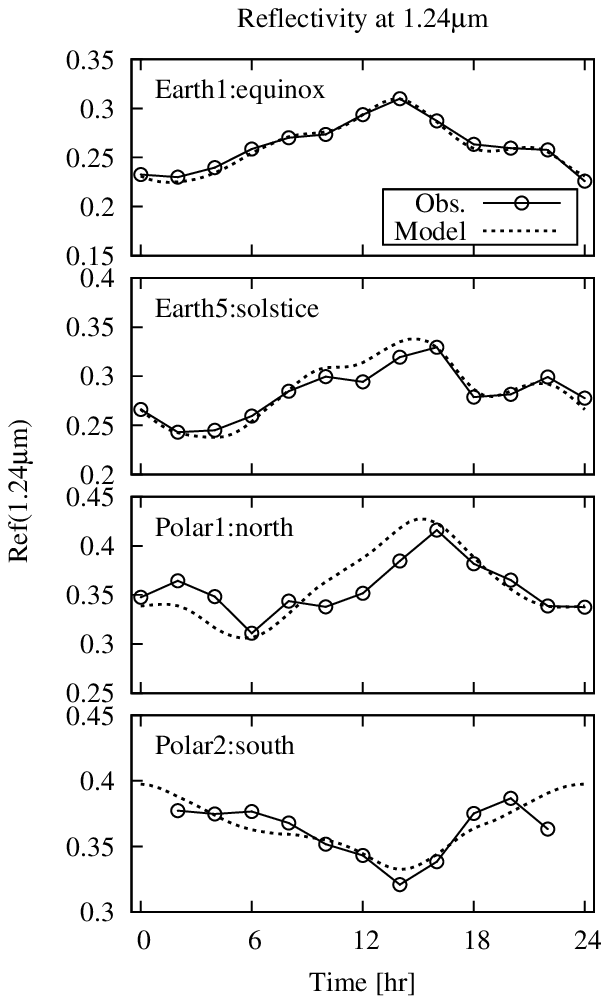}
     \end{center}
\end{minipage}
   \begin{minipage}{0.33\hsize}
     \begin{center}
       \includegraphics[width=\hsize]{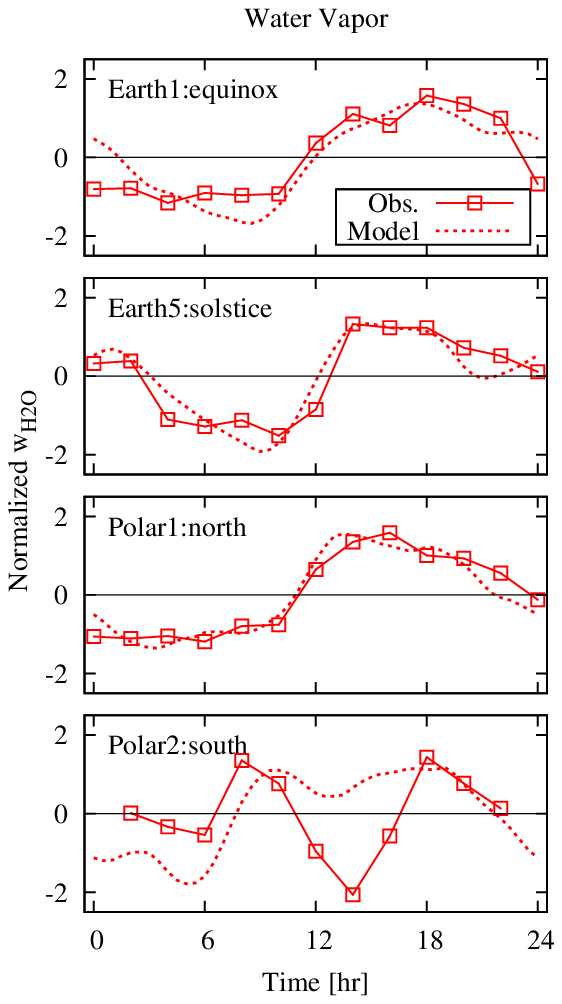}
     \end{center}
\end{minipage}
   \begin{minipage}{0.33\hsize}
     \begin{center}
       \includegraphics[width=\hsize]{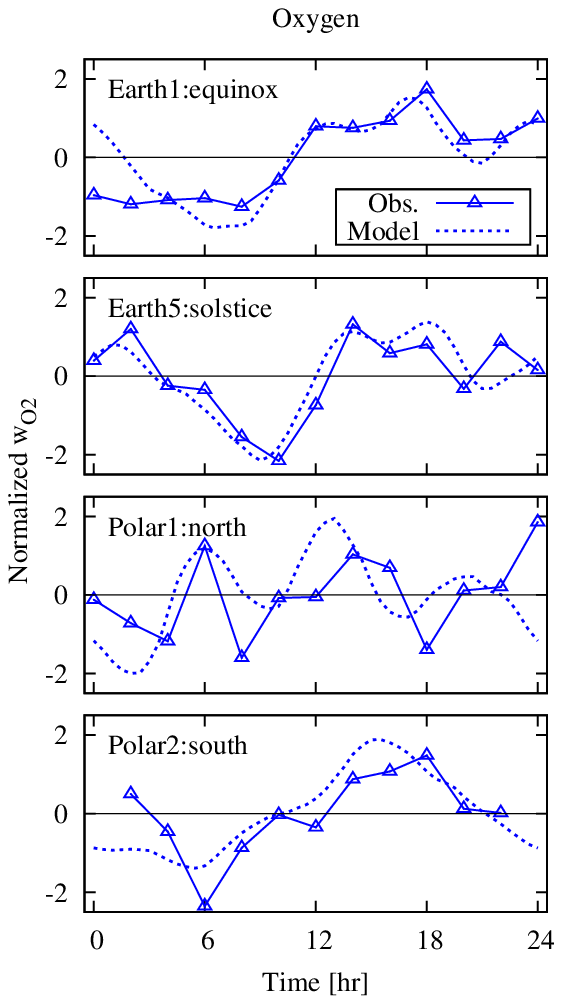}
     \end{center}
\end{minipage}
\caption{Comparison between the EPOXI data (symbols) and simulations
(dashed lines) in terms of continuum level measured at 1.24$\mu $m
(left), water vapor absorption (middle), and oxygen absorption
(right). } \label{fig:obs-model}
\end{figure*}

Figure \ref{fig:obs-model} compares the EPOXI data (symbols) to our
model predictions (dotted lines).  The left, middle and right panels
correspond to the disk-integrated continuum level, H$_2$O absorption,
and O$_2$ absorption, respectively.  The anisotropic parameter for bulk
cloud scattering $\alpha $ in equation (\ref{eq:2-stream}) is determined
so that the disk-averaged continuum level matches the observed value
(Table \ref{tab:norm}).

\begin{table}[h]
\begin{center}
\caption{Normalization Parameters and Comparison in Daily Average and 
Standard Deviation between Simulations and Observations. } 
\label{tab:norm}
\begin{tabular}{c|cccc} \hline \hline
& Earth1 & Earth5 & Polar1 & Polar2 \\ \hline
$\alpha $ %
&0.61        &0.64        &0.73        &0.79        \\
$\bar w_{\rm H2O,model}/\bar w_{\rm H2O,obs}$%
&0.85         &0.87        &0.88        &0.88        \\
$\bar w_{\rm O2,model}/\bar w_{\rm O2,obs}$%
&0.65        &0.66        &0.64        &0.63        \\
$\sigma _{\rm H2O,model}/\sigma _{\rm H2O,obs}$%
&1.32        &1.73        &0.70        &0.92        \\
$\sigma _{\rm O2,model}/\sigma _{\rm O2,obs}$        %
&0.57        &0.77        &0.61        &0.88        \\ \hline
\end{tabular}
\end{center}
\end{table}

We first consider the comparison between the simulation and the 
observation in terms of the daily average, $\bar w$, and the standard 
deviation, $\bar \sigma $.
Table \ref{tab:norm} summarizes the ratio of the daily average of the 
simulated $\bar w_{\rm model}$, to the observed one, $\bar w_{\rm obs}$, 
$\bar w_{\rm model}/\bar w_{\rm obs}$, and that of standard deviation 
$\bar \sigma _{\rm model}/\bar \sigma _{\rm obs}$.
The daily simulation average typically results in a 10-15\% 
underestimation for H$_2$O and 35\% underestimation for O$_2$. The 
underestimation is likely due to our ``opaque cloud'' assumption which 
tends to diminish the absorption depth; in reality cloud cover does not 
completely prevent the light from going through the lower atmosphere.
Uncertainty can also come from the difficulty in determining the continuum 
level because the absorption bands we are considering are surrounded by 
other absorption features.

The variation pattern, which is of our primary interest in this paper, is 
less sensitive to those uncertainties.
The absorption depths normalized so that the daily average is 0 and the 
standard deviation is unity are exhibited in the middle and right panels 
of Figure \ref{fig:obs-model}.
In most cases, the overall variation patterns of both H$_2$O and O$_2$ are 
well reproduced by our opaque cloud model.  In particular, the equatorial 
data (Earth1, Earth5) exhibit striking agreement with the model 
predictions.
For polar observations, the agreement is somewhat degraded, especially for 
H$_2$O
variations in the Polar2 data.  This may be partly ascribed to the fact 
that
there is a slight mismatch between the time of the EPOXI observation and
that of the input atmospheric data of our simulation; the local time of 
MODIS observation is 10:30am for {\it Terra} and 12:10pm for {\it Aqua}, 
while the disk-integrated spectra observed by  EPOXI reflects the 
information of slices with different local times.
In addition, the assumed vertical profile for H$_2$O
(the US standard atmosphere) is not expected to be as accurate for the 
equatorial/polar regions.
While our model is admittedly rough in this regard, it is
encouraging that such a simple model reproduces the
general trends and features observed by EPOXI fairly well.

\section{Different Variation Patterns of H$_2$O and O$_2$ Absorption Depths}
\label{s:dis}

We now focus on the intriguing differences in behavior between H$_2$O and 
O$_2$.
In the Earth1 and Earth5 data, for instance, note the small deviation of 
the H$_2$O pattern from the O$_2$ pattern, including the sharper bump of 
$w_{\rm O2}$ at $t=18$[hr] than $w_{\rm H2O}$.
Polar1 observations show the variation of $w_{\rm H2O}$ rising at $11\le t 
{\rm [hr]} \le 19$, which is not present in the variation of $w_{\rm O2}$. 
In addition, only $w_{\rm O2}$ has a local peak at $t=6$.
Significantly and reassuringly, these divergences are reproduced by our 
simulations, which allows us to isolate the origin of this difference.

\begin{figure}[!h]
   \begin{center}
     \includegraphics[width=0.9\hsize]{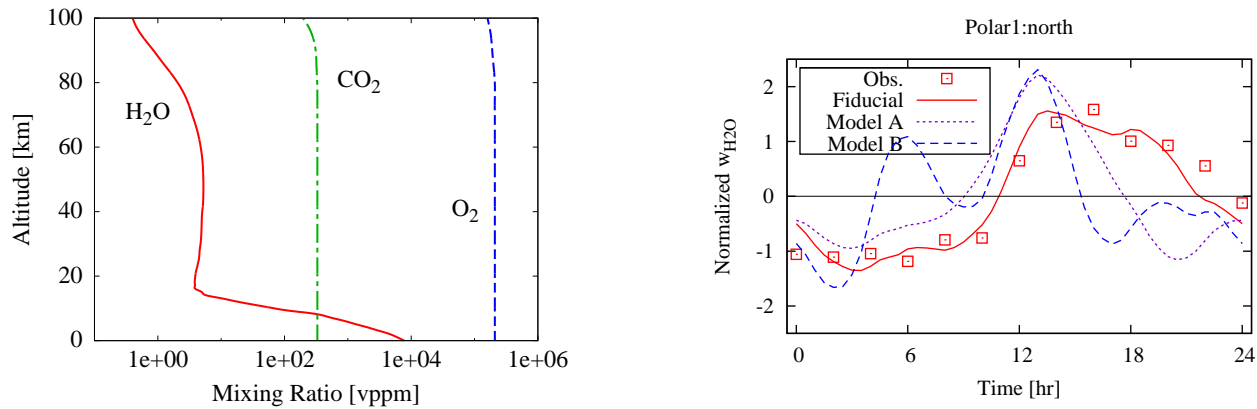}
   \end{center}
\caption{Vertical profiles of H$_2$O (red solid) and O$_2$ (blue dashed)
taken from US standard model. } \label{fig:profile}
\end{figure}

\begin{figure}[!ht]
     \begin{center}
       \includegraphics[width=0.9\hsize]{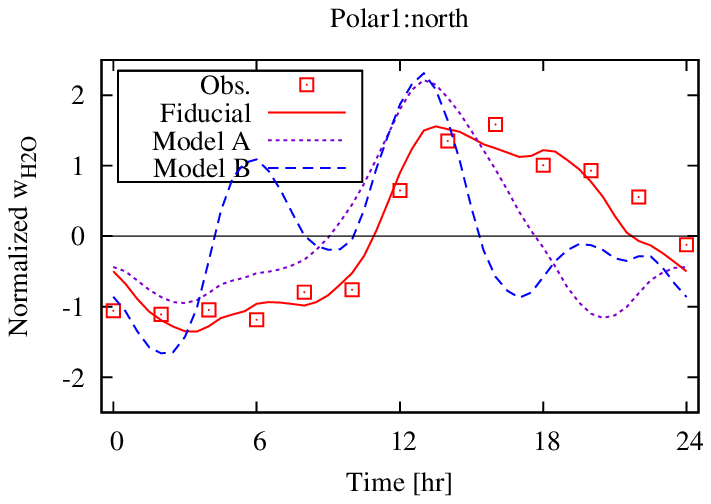}
     \end{center}
\caption{Comparison in H$_2$O band variation between observation and
simulations with different assumption for water vapor
distribution. Model A adopts the water vapor vertical profile of US 
standard model but neglect the horizontal inhomogeneity. Model B assumes 
constant mixing ratio in the troposphere as is the case for O$_2$. } 
\label{fig:NS_comp}
\end{figure}

Let us discuss the possible causes of the different diurnal pattern of
H$_2$O and O$_2$ bands.  Given that the absorption depth in our model is
directly linked to the total number of molecules above the cloud layers, 
the
different behaviors of H$_2$O and O$_2$ exhibited in Figure
\ref{fig:obs-model} can only be ascribed to the differences in their 
column
number density distribution in the planetary atmosphere.

In the case of the Earth, the distribution of water vapor differs from
that of O$_2$ in at least two ways.  Firstly, the total column density
of atmospheric water vapor is, unlike O$_2$, a strong function of
latitude (highly concentrated in the equatorial region) with additional
local/temporal fluctuations.  Secondly, the vertical profile of water
vapor mixing ratio typically decreases as a function of altitude due to
condensation, in contrast to the constant mixing ratio (volume fraction
of the molecule) of O$_2$ in the troposphere (Figure
\ref{fig:profile}).

In order to evaluate these effects quantitatively, we run simulations of
H$_2$O band variation with different assumptions for its atmospheric
distribution.
We consider two model distributions in addition to our fiducial model 
described in Section \ref{ss:model}: Model A adopts the water vapor 
vertical profile of the US
Standard Model everywhere, and thus the mixing ratio decreases with
altitude (red solid line in Figure \ref{fig:profile}). Model B instead 
assumes a constant mixing ratio of H$_2$O in the troposphere as is the 
case for O$_2$ (blue dashed line in Figure \ref{fig:profile}).  Both 
Models A and B neglect the inhomogeneity of water vapor over the surface 
of the planet.

These three simulations and observations in the Polar1:north case are 
plotted in Figure \ref{fig:NS_comp}.  The observed
variation of H$_2$O agrees well with fiducial model, while the
other two models that neglect the horizontal or surface inhomogeneity 
and/or
peculiar vertical profiles fail to match the observed data.  We also
note that the Model B (blue line in Figure \ref{fig:NS_comp}) prediction 
of
$w_{\rm H_2O}$ is very close to that of the simulation for O$_2$ in the
Polar1 case (solid line in the third panel of O$_2$ in Figure
\ref{fig:obs-model}). This indeed implies that that the differences in 
variation
patterns basically reflect their differing spatial distributions in the 
atmosphere.

Both of the above-mentioned properties of water vapor distribution are
consequences of the coexistence of multiple phases of water on Earth.  The
horizontal inhomogeneity of water vapor originates from the temperature
gradient over the surface (since the vaporization rate is basically
determined by the surface temperature) and/or the inhomogeneous
distribution of liquid/solid reservoirs on the surface.
The water vapor distribution is not mixed well since the mean
residence time (MRT) of atmospheric H$_2$O ($\sim $10 days) is much
shorter than the time scale of the vertical and global mixing times of the 
atmosphere
(both $\sim $ 1 year). In contrast, the MRT of O$_2$ is $\sim $4000 years,
thus insuring that O$_2$ is well mixed and only very slowly changing, if 
at all.

The short MRT of water vapor implies that there is a non-negligible flux 
of
H$_2$O into/out of atmosphere and the existence of a substantial water 
reservoir
(compared to the abundance in the atmosphere), such as the oceans on the
Earth.  Figure \ref{fig:NS_comp} also indicates that the difference in the
vertical profile alone may lead to differences in the variation
of absorption bands, but this is also related to the condensation of water 
in the atmosphere.  In either case, the deviation of absorption features 
of
H$_2$O from the well-mixed gas like O$_2$ may serve
as an indicator of phases transitions happening in the atmosphere or on
the surface of a planet.
This signature of the Earth's ``water cycle'' could be an observable
indicator of a similar climate on a terrestrial exoplanet.

\section{Summary and Discussion}
\label{s:sum}

In this paper we examined the diurnal variability of the absorption
depths of the two important molecules, H$_2$O and O$_2$, in the
disk-integrated spectra of the Earth.  We analyzed the H$_2$O band
centered at $\sim 1.13\mu $m and O$_2$ band centered at $\sim 1.27\mu $m
observed by EPOXI and found that these absorption bands show diurnal
variations correlated with uneven cloud cover.  A simple opaque cloud
model, which assumes that the cloud completely blocks the atmospheric 
signatures
below the cloud top layer, is able to reproduce the basic variation
patterns of the H$_2$O and O$_2$ bands using pixel-level cloud data
obtained with Earth Observing Satellite.  Thus we conclude that the
non-uniform cloud cover distribution dominates the observed
diurnal variations of those molecular absorption depths.

However, we also found that the diurnal variability patterns of H$_2$O and
O$_2$ bands are not identical, and the differences
originate from the inhomogeneous distribution of water vapor in the
atmosphere.  The variability pattern of O$_2$ is
basically explained by the the cloud cover distribution because it is
well mixed in the troposphere of the Earth. In contrast, the variability
pattern of H$_2$O is well reproduced only with additional information on 
the
vertical profile and spatial inhomogeneity of atmospheric water vapor as a 
model input.
The nature of the water vapor distribution in the atmosphere is linked
to the fact that H$_2$O circulates in the planetary surface layer
changing its phase among water vapor, liquid water
(ocean/lake/pond/river) and/or water ice.  Therefore, different
behavior in the variability patterns of O$_2$ and H$_2$O absorptions may
carry information on the inhomogeneous phase distribution of H$_2$O
in the surface layer.

Our study of the Earth demonstrates the possible role of the variability 
of absorption bands in characterizing the surface environments of Earth-analogs.  
If future instruments eventually succeed in detecting scattered 
light from such exoplanets, their
time-resolved measurements of the absorption bands will reveal the
uneven cloud cover and/or inhomogeneous spatial distribution of 
atmospheric constituents.  Our
current results suggest that the comparison in variation pattern between
presumably well-mixed gases (such as O$_2$) and H$_2$O may indicate 
frequent phase transitions
of water in the surface layer and thus serve as a probe of habitability,
in a complementary fashion to the direct detection of liquid water
\citep[e.g.][]{williams2008,oakley2009,robinson2010,zugger2010} and the 
search for other
potential biosignatures
\citep[e.g.][]{kaltenegger2010,kawahara2012}.

While we focused on absorption bands in the NIR range where the EPOXI
data are available, it is natural to expect that other bands also
exhibit similar variation patterns at wavelengths dominated by scattering 
of the primary
star's light, rather than planetary thermal emission.
In the visible range, there are O$_2$ bands at 0.69$\mu $m (equivalent
width $\sim $13.5\AA) and at 0.76$\mu $m (equivalent width $\sim $47.8\AA)
\citep[e.g.][]{palle2009} as well as several H$_2$O bands.

Inhomogeneity of an atmospheric constituent becomes appreciable when its
MRT is short compared to the time-scale of the atmosphere mixing. This
is equivalent to there being a substantial flux of the molecular species 
in question
between the atmosphere and some surface reservoir, i.e., substantial 
compared
to the total amount residing in the atmosphere.  Besides
vaporization, photochemical production (e.g. O$_3$) or biotic production
(e.g. N$_2$O, CH$_4$) could lead to such a situation.  If sufficiently 
accurate data
become available in the future, we may be able to interpret implied
inhomogeneous distributions in terms of those or other specific 
mechanisms.

\acknowledgments

We thank an anonymous referee for constructive comments. 
We are grateful to Timothy A. Livengood for his kind assistance in
obtaining EPOXI data.  We acknowledge support from the Global
Collaborative Research Fund (GCRF) ``A World-wide Investigation of Other
Worlds'' grant and the Global Scholars Program of Princeton University.
Y.F. is supported by JSPS (Japan Society for the Promotion of Science)
Fellowship for Research, DC:23-6070.  The work of Y.S. is supported in
part from the Grant-in-Aid Nos. 20340041 and 24340035 by JSPS.  E.L.T.
was supported in part by the World Premier International Research Center
Initiative (WPI Initiative), MEXT, Japan.



{\appendix
\section{Variation of CO$_2$}
\label{ap:CO2}

We also examine the variability of CO$_2$, another absorption feature 
imprinted in the observed spectra.
Absorption bands of CO$_2$ exist around 1.6$\mu $m and 2$\mu $m (Figure 
\ref{fig:spec}).
These CO$_2$ bands override the wings of H$_2$O absorption, making it 
harder to determine the continuum level.
We pick up the absorption band at 1.59-1.62$\mu $m, which is likely to be 
least affected by H$_2$O absorption.
The peak-to-throat variation amplitude of four observations are 
5-20\%\footnote{%
The large peak-to-throat amplitude of Earth1:equinox is at least partly 
due to the failure in continuum determination. In Earth1:equinox only, the 
observed wavelength grid changes for each exposure and the sharpness of 
the very narrow feature at 1.58$\mu m$, which is critical to determine the 
continuum level, also changes. This results in the uncertainty in 
estimation of equivalent width. %
}(Table \ref{tab:var_CO2}), on the same level of O$_2$ and H$_2$O.

Figure \ref{fig:obs-model_CO2} shows the diurnal variation patterns of 
CO$_2$ band at 1.59-1.62$\mu $m extracted from EPOXI data as well as those 
of our simulation following the same procedure as H$_2$O/O$_2$ and 
assuming the uniform distribution (the assumed growth curve of this band 
is displayed in Figure \ref{fig:cog}).
The normalization factors are summarized in Table \ref{tab:norm_CO2}.
We see the clear similarity in variation patterns to O$_2$ rather than 
H$_2$O. For instance, the clearer bumps at $t=18$ of Earth1 and Earth5 
observations and the gradient in Polar2 observation.
This result is consistent with our discussion in Section \ref{s:dis}, 
since the MRT of CO$_2$ is 3-5 years and slightly longer than the 
tropospheric mixing timescale.
The behavior of Polar1 is less conclusive and the observation does not 
match the simulation so well . Although we do not identify the cause, it 
might be related to some level of inhomogeneity of CO$_2$ due to the 
intermediate MRT.

\begin{table}[!htdp]
\begin{center}
\caption{Diurnal Variation of Continuum Level, Equivalent Width of CO$_2$ 
at 1.6$\mu $m. } \label{tab:var_CO2}
\begin{tabular}{l|cc} \hline \hline
& \multicolumn{2}{c}{CO$_2$ at 1.6$\mu $m}\\
&ave.[\AA]        & (max-min)/ave \\ \hline
Earth1:equinox (Mar.2008)         &12.1        &80.1\%\\
Earth5:solstice (Jun.2008)        &17.8        &17.4\%\\
Polar1:north (Mar.2009)                &17.4        &11.3\%\\
Polar2:south (Oct.2009)                &17.9        &5.4\%\\ \hline
\end{tabular}
\end{center}
\end{table}%

\begin{figure}[!h]
     \begin{center}
       \includegraphics[width=60mm]{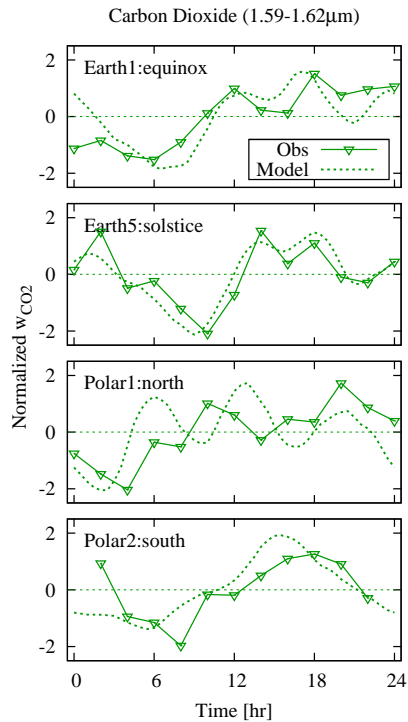}
     \end{center}
\caption{Diurnal Variations of CO$_2$ absorption bands around 1.59-1.62 
$\mu$m. EPOXI data (symbols) and simulations (dashed lines) are compared. 
} \label{fig:obs-model_CO2}
\end{figure}

\begin{table}[!htdp]
\begin{center}
\caption{Normalization Parameters and Comparison in Daily Average and 
Standard Deviation between Simulations and Observations. } 
\label{tab:norm_CO2}
\begin{tabular}{c|cccc} \hline \hline
& Earth1 & Earth5 & Polar1 & Polar2 \\ \hline
$\bar w_{\rm CO2,model}/\bar w_{\rm CO2,obs}$%
&1.14         &0.85        &0.95        &0.90        \\
$\sigma _{\rm CO2,model}/\sigma _{\rm CO2,obs}$        %
&0.23        &1.23        &0.74        &1.75        \\ \hline
\end{tabular}
\end{center}
\end{table}

}

\end{document}